\documentclass[aps,prl,reprint,superscriptaddress]{revtex4-1}

\usepackage{graphicx,color}
\usepackage{subfig}
\usepackage{amsmath} 
\usepackage{verbatim} 
\usepackage{mathrsfs}
\usepackage{amsfonts}
\usepackage[english]{babel}
\usepackage[utf8]{inputenc}
\usepackage[T1]{fontenc}
\usepackage{hyperref}

\newcommand{\nn}{\nonumber \\}
\newcommand{\bra}{\left\langle}
\newcommand{\ket}{\right\rangle}
\newcommand{\beq}{\begin{equation}}
\newcommand{\eeq}{\end{equation}}
\newcommand{\bqr}{\begin{eqnarray}}
\newcommand{\eqr}{\end{eqnarray}}

\newcommand{\bii}{\begin{itemize}}
\newcommand{\eii}{\end{itemize}}

\def \p{\partial}
\def \d{\delta}

\def \({\left(}
\def \){\right)}
\def \[{\left[}
\def \]{\right]}

\begin{document}

\author{Gabriel León}
\email{gleon@df.uba.ar}
\affiliation{Departamento de Física, Facultad de Ciencias Exactas y Naturales, Universidad de Buenos Aires, Ciudad Universitaria -
 Pab. I, 1428 Buenos Aires, Argentina.}
\affiliation{Grupo de Astrof\'isica, Relatividad y 
Cosmolog\'ia, Facultad de Ciencias 
Astron\'omicas y Geof\'isicas, Universidad Nacional de La Plata, Paseo del Bosque S/N 
(1900) La Plata, Argentina}
\author{Abhishek Majhi}
\email{abhishek.majhi@gmail.com}
\affiliation{Instituto de Ciencias Nucleares, Universidad Nacional Aut\'onoma de M\'exico, Mexico City, 04510, Mexico.}
\author{Elias Okon}
\email{eokon@filosoficas.unam.mx}
\affiliation{Instituto de Investigaciones Filos\'oficas, Universidad Nacional Aut\'onoma 
de M\'exico, Mexico City, 04510, Mexico.}
\author{Daniel Sudarsky}
\email{sudarsky@nucleares.unam.mx}
\affiliation{Instituto de Ciencias Nucleares, Universidad Nacional Aut\'onoma de 
M\'exico, Mexico City, 04510, Mexico.}

\title{Reassessing the link between B-modes and inflation}

\begin{abstract}
We reevaluate the predictions of inflation regarding primordial gravity waves, which should appear as B-modes in the CMB, in light of the fact that the standard inflationary paradigm is unable to account for the transition from an initially symmetric state into a non-symmetric outcome. We show that the incorporation of an element capable of explaining such a transition dramatically alters the prediction for the shape and size of the B-mode spectrum. In particular, we find that by adapting a realistic objective collapse model to the situation at hand, the B-mode spectrum gets strongly suppressed with respect to the standard prediction. We conclude that the failure to detect B-modes in the CMB does not rule-out the simplest inflationary models.
\end{abstract}

\maketitle


The exquisite matching between observations of temperature anisotropies in the CMB and the generic predictions of inflation provides a powerful justification for the recent consolidation of the inflationary paradigm as a cornerstone of modern cosmology. However, such a paradigm also makes predictions for the shape and amplitude of primordial gravity waves, which should be observable in the B-mode polarization of the CMB. The fact that, so far, such B-modes have not been observed, has been used to severely constrain the set of viable inflationary models \cite{nullB-modes1,nullB-modes2,Kamionkowski}. Moreover, the recent detection of gravity waves by LIGO \cite{LIGO} removes all lingering doubts about the reality of such waves (if there were any remaining, given the dramatic indirect evidence provided by the Binary Pulsar studies \cite{BinPul}). This only increases the urgency to resolve the tension generated by the predictions regarding B-modes and our failure to detect them. 

As initially argued in \cite{Seeds}, the standard inflationary account for the emergence of the primordial perturbations lacks a crucial element capable of accounting for the transition from the initially homogeneous and isotropic quantum state into a state lacking such symmetries. The objective of this letter is to show that the adoption of a framework that explicitly introduces this element drastically modifies the predictions for the shape and amplitude of the primordial gravity waves (without also altering confirmed predictions of inflation)\footnote{We note that a calculation analogous to the present one, made in terms of the very simple collapse toy model used in \cite{Seeds} rather than the full fledged CSL model, has been presented in \cite{BModesLeon}.}. As a result, we conclude that the impact of the B-polarization null-results on the viability of various inflationary models has to be reexamined. In particular, we establish that, contrary to what has been argued, the failure to detect B-modes in the CMB does not rule-out the simplest inflationary models.

The crucial observation overlooked by the standard account is concerned with the way in which the \emph{quantum fluctuations} of the completely homogeneous and isotropic state that characterizes the early stages of inflation (i.e., the Bunch-Davies vacuum or related states), are supposed to transform into actual inhomogeneities and anisotropies. The problem is that such ``fluctuations'' are nothing more than a characterization of the \emph{width} of the wave function for the corresponding field degrees of freedom and not acutal, physical fluctuations. Therefore, they do not represent deviations from isotropy and homogeneity and cannot be used to explain departures from such symmetries. Of course, all these issues are connected with the notorious conceptual difficulties of quantum theory \cite{MP1,MP2,MP3} and, as such, they are often regarded as having no physical, only ``philosophical'' implications. However, the situation under consideration is one of the few clear instances where such expectations do not hold. As we will see below in detail, the viability of many inflationary models, including some of the simplest ones, strongly depends on the point of view adopted regarding interpretational issues.
 
 
Before advancing our own proposal, we review a few aspects of the standard approach to inflationary cosmology. The starting point is the action of a scalar field $\phi$, the inflaton, with potential $V$, coupled to gravity
\beq
\label{eq_action}
S=\int d^4x \sqrt{-g} \left[ {\frac{1} {16\pi G}} R[g] - 1/2\nabla_a\phi
\nabla_b\phi g^{ab} - V(\phi)\right].
\eeq
One then splits both the metric $g$ and the scalar field $\phi$ into spatially homogeneous 
backgrounds, $g_0$ and $\phi_0$, and inhomogeneous fluctuations $\delta g$ and 
$\delta\phi$. The backgrounds are taken to be the spatially flat FRW universe, with line 
element $ ds^2=a(\eta)^2\left[- d\eta^2 + \delta_{ij} dx^idx^j\right]$, and an homogeneous 
scalar field $\phi_0(\eta)$ (with $\eta$ a conformal time). The solution for the scale 
factor $a(\eta)$ during the inflationary era is $a(\eta)=[-1/H \eta]^{(1+\epsilon)}$, with 
$ H ^2\approx (8\pi/3) G V$ and $\epsilon \equiv 1-\dot{\mathcal{H}}/\mathcal{H}^2$ (where 
$\mathcal{H} \equiv \frac{\dot a}{a} $). The parameter $\epsilon$ is called the slow-roll 
parameter and during the inflationary stage is considered to be very small, $\epsilon \ll 
1$. Also, the scalar field $\phi_0$ is supposed to be in the slow-roll regime, i.e., 
$\dot\phi_0= - (a^3/3 \dot a)V'$. We ignore the fact that the functional form of $a(\eta)$ 
changes after inflation and, for definiteness, we set $a(\eta)=1$ at the present 
cosmological time and assume that the inflationary regime starts at $\eta = -T$ (note that $T>0$) and ends 
at $\eta=\eta_0$, which is negative and very small in absolute terms.

Regarding the perturbations, one works in the, so-called, longitudinal gauge, where the perturbed metric is written as
\begin{equation}
ds^2=a(\eta)^2\left\{ -(1+ 2 \Psi) d\eta^2 + \left[ (1- 2\Psi)\delta_{ij} + h_{ij} \right] dx^idx^j\right\},
\end{equation}
where $\Psi$ stands for the scalar perturbation, usually known as the \emph{Newtonian potential}, and the transverse traceless $h_{ij}$ stands for the tensor perturbations, corresponding to gravity waves. 

The next step in the standard treatment is to introduce quantum mechanics into the picture. This is done by quantizing the perturbations $\Psi$, $\delta\phi$ and $h_{ij}$ as fields on the background provided by the spacetime $g_0$ and the classical inflaton field $\phi_0$. In fact, the equations 
of motion for these perturbations lead to an interaction between $\Psi$ and $\delta\phi$ that makes 
it convenient to directly quantize the field $v =a ( \delta\phi + \frac {\dot \phi_0}{\mathcal{H}} \Psi ) $, known as the Mukhanov-Sasaki variable. One also assumes that the quantum state of the field is the Bunch-Davies vacuum \footnote{Or some other appropriate vacuum state characterized, in the asymptotic past, by initial mode conditions taken in analogy with the Minkowski vacuum. The point, though, is that such states are equivalent to the Bunch-Davies one regarding the homogeneity and isotropy properties.}. Finally, one identifies the vacuum uncertainties in the spatial Fourier components of $v$, $\Psi$ and $h_{ij}$ with actual, physical inhomogeneities in the CMB, leading, as is well-known, to a spectacular matching between predictions and observations.

Now, since both the scalar and tensor perturbations are treated equally in this setting, it is not surprising, given a specific inflationary model, for there to be a close similarity in their predicted amplitudes and shapes. In fact, the standard estimates for the power spectra of the scalar and tensor perturbations are given by
\begin{equation}
P^S_s (q) \simeq \frac {G H^2}{q^3\epsilon} \quad \text{and} \quad
P^S_h (q) \simeq \frac {G H^2}{q^3}
\label{usual-exp}
\end{equation}
respectively, where the superscript $S$ stands for `standard' scenario (in contrast with the `collapse' scenario introduced below) \footnote{See \cite{BmodesAlt1,BmodesAlt2} for alternative derivations of the tensor spectrum within the standard approach.}. Therefore, for $r$, the ratio of the amplitudes in tensor and scalar perturbations, one finds $r \simeq \epsilon$. As $\epsilon $ is also connected with the tilt in the scalar spectrum \cite{Tilt}, empirical constraints on that quantity can be used to limit the viability of various simple inflationary models. In particular, measurements of the amplitude of scalar perturbations, combined with the failure to detect tensor perturbations, have been used to discriminate among inflationary models and even to rule-out some of the simplest ones \cite{nullB-modes1,nullB-modes2,Kamionkowski}.

There are, however, a few problematic aspects within the standard approach that call into question some of its conclusions. The first, already mentioned above, is the fact that the identification between quantum uncertainties and actual inhomogeneities is not well-founded. The issue, again, is that such uncertainties represent the width of the wave function and not actual fluctuations. Therefore, just as the spread in position in the ground state of a 1D harmonic oscillator does not imply a breakdown of the reflection symmetry,
the fluctuations in the vacuum state of the inflaton and metric perturbations should not be read as implying a breakdown of the translation and rotation symmetries of such a state.

The upshot is that the standard approach is incapable of explaining the emergence of structure out of an initially homogeneous and isotropic state (see \cite{shortcomings} for a detailed discussion of the issue). This point has been explicitly acknowledged in various textbooks on the subject, \cite{Weinberg, Muckanov}, and has been the object of several papers \footnote{It is fair to say that many cosmologists do not fully agree with this point of view, even though an increasing number accepts that there are, in fact, various unsolved issues in this regard (see, e.g., \cite{Jerome,TP}).}. Moreover, in \cite{Multiple,CSL-Inflation,SSC,Amplitude-Inflation} we have shown that the adoption of objective collapse models, developed in connection with foundational issues within quantum theory (see \cite{collapse} for an extensive review of these models),
allows for a straightforward resolution of the issue at hand \footnote{In \cite{Okon:2013lsa,OSBH,Okon-New} we argue that objective collapse models also offer attractive benefits in connection many with other open problems within contemporary physics.}. Below we explore the consequences of adopting such models in the analysis of the inflationary primordial gravity waves.

A second potential problem within the standard approach has to do with the fact that, as we saw, both the metric and the field backgrounds are treated classically and both the metric and inflaton field perturbations are quantized. However, it is not clear that quantizing the metric perturbation is the correct thing to do. There is of course indisputable evidence of the quantum nature of matter, from which it follows that a theory of gravity that, unlike general relativity, acknowledges the quantum character of matter, is necessary. But that does not necessarily mean that what we need is a theory in which the metric itself is quantized. 
The prevailing perception is that combining a classical theory of gravity with a quantum theory of matter would be inconsistent and, in fact, there are a number of arguments in the literature against the viability of a half-and-half theory. Nevertheless, none of those arguments is really conclusive \cite{CH,Mat1,Mat2,Carlip}. 

Moreover, even if we were to agree that, at the fundamental level, gravitation itself is quantum mechanical in nature, that does not automatically mean that the metric degrees of freedom are the ones that need to be treated quantum mechanically. In fact, over the years, various arguments questioning whether the spacetime metric should be treated quantum mechanically have been put forward (see for instance \cite{Jacobson,Mat1,Hu}). The idea is that the spacetime geometry might emerge from deeper, non-geometrical and fundamentally quantum mechanical degrees of freedom (see, e.g., \cite{Em1,Em2,Em3,Em4,Em5}); and, just as one does not directly quantize macroscopic, thermodynamic variables or the degrees of freedom of, say, the heat equation, one would not think of quantizing the metric if it were non-fundamental.

Finally, as it is well-known, canonical approaches to quantization generically lead to timeless theories (see, e.g., \cite{Time-in-QG}) in which the recovery of fully covariant spacetime notions becomes, by itself, a nontrivial task, often only achieved in an approximated sense while applying a semiclassical treatment. Therefore, even if in the deep quantum gravity regime the space-time degrees of freedom are fully quantum mechanical, it does not seem unreasonable to imagine that in order to correctly describe scenarios in which \emph{temporal} processes, such as inflation, are taking place, one needs to rely on a semiclassical treatment. 


With all these ideas in mind, the starting point of our analysis of the inflationary perturbations is the so-called semiclassical Einstein equation, 
\begin{equation}
\label{SCE}
R_{\mu\nu} -(1/2) g_{\mu\nu} R =8\pi G \langle \hat {T}_{\mu\nu} \rangle ,
\end{equation}
which treats gravitation classically and all other fields quantum mechanically \footnote{See \cite{BModes-class1,BModes-class2} for alternative analyses of the tensor power spectrum within a semi-classical approach.}. We assume such a framework to be a valid approximation at the regime under consideration, which is well after the full quantum gravity regime has been left behind.

Next, in accordance with the standard inflationary account, we assume that the state of the universe before the time at which the seeds of structure emerge is given by the homogeneous and isotropic Bunch-Davies vacuum and the homogeneous and isotropic classical FRW spacetime. However, inspired by the dynamical reduction program, we assume the state of the matter field to undergo spontaneous and stochastic quantum collapses into states which need not share the symmetries of the Bunch-Davies vacuum. Since, as a result, $ \langle {T}_{\mu\nu} \rangle $ for such states will not have the symmetries of the initial state, the semiclassical Einstein equation will lead to a geometry that, generically, will no longer be homogeneous and isotropic.

To implement the spontaneous collapses we employ the CSL model which, in the non-relativistic context, is characterized by a modified evolution of the quantum state given by
\begin{equation}
|\psi,t\rangle_w={\cal T}e^{-\int_{0}^{t}dt'\big[i\hat H+\frac{1}{4\lambda_0}[w(t')-2\lambda_0\hat A]^{2}\big]}|\psi,0\rangle ,
\end{equation}
where ${\cal T}$ is the time-ordering operator, $ \hat H$ the usual Hamiltonian, $\lambda_0$ a new fundamental parameter, $w(t)$ a random classical function of time and $\hat A$ the so-called collapse operator (in the non-relativistic case, such operator is taken to be a smeared position operator for each particle). The second component of the CSL model is a suitable probability rule for the realization of the stochastic function $w(t)$ (see \cite{Pearle:89} for details). In the context at hand, the theory has to be generalized to the case of infinite degrees of freedom and one has to construct a suitable collpase operator out of the inflaton field (this is the content of \cite{CSL-Inflation}; see also \cite{weight} for a general discussion of objective collapse models in cosmological scenarios).

Going back to inflation, regarding the scalar perturbations, the semiclassical Einstein equation leads to the following constraint equation for the Newtonian potential
\begin{equation}
 \nabla^2 \Psi =4\pi G \dot \phi_0 \langle \hat{\delta \dot\phi} \rangle , 
\label{main3}
\end{equation}
where we have made use of the specific form of the scale factor during inflation in the slow role regime. 
It is then clear that, as soon as $\langle \hat{\delta \dot\phi} \rangle$ deviates from zero, so 
will the Newtonian Potential. In order to derive concrete predictions, one considers an ensemble of universes, all with the Bunch-Davies vacuum as the initial state for the inflaton perturbation, then calculates the CSL evolution of each of them and computes an ensemble average over possible stochastic realizations (see \cite{CSL-Inflation} for details). The result is a prediction for the scalar power spectrum which, after taking into account the modifications occurring in the post reheating regime, \cite{Amplitude-Inflation}, agrees with the usual estimate for the scalar perturbation power spectrum of Eq. \ref{usual-exp}.

Regarding the tensor perturbations, the equation of motion analogous to \ref{main3} is given by
\begin{equation}
(\p^2_0-\nabla^2)h_{ij}+2 (\dot a/a)\dot h_{ij} =16\pi G \langle (\p_{i} 
\hat{\d\phi})\rangle\langle(\p_{j} \hat{\d\phi}) \rangle ^{\textrm{tr-tr}},
\label{gw}
\end{equation}
where the superscript tr-tr stands for the transverse traceless part of the expression. 
Note as before that, even though (the renormalized value for) $\langle (\p_{i} 
\hat{\d\phi})\rangle\langle(\p_{j}\hat{\d\phi})\rangle $ vanishes when evaluated in the 
Bunch-Davies vacuum, it will become non-vanishing in the quantum state of the field that results from the spontaneous collapses.

Now, from the above expressions it is easy to see the basic difference, within this scheme, between the scalar and tensor metric perturbations. While the former are seeded by linear terms in perturbations of the scalar field (see Eq. \ref{main3}), the latter are seeded by quadratic terms in such perturbations (see Eq. \ref{gw}). This difference represents a radical departure from the standard approach with crucial consequences: as long as one is in any regime where perturbation theory makes sense, the second spectrum will be much smaller than the first. 

Going back to Eq. \ref{gw}, taking (without loss of generality) the wave propagation direction to be $\vec q = q \hat 
z$, orienting the polarization components along the $ x, y $ directions and dropping the 
indices, the solution for the Fourier transform of $h(\vec x,\eta)$ is given by
\begin{eqnarray}
h(\vec q,\eta) &=& -i h^+(\vec q,\eta) \int_{-T}^\eta d \eta_1 \frac{h^-(\vec q,\eta_1) 
 S(\vec q, \eta_1) }{H^2 \eta_1^2} \nn
& &+ i h^-(\vec q,\eta) \int_{-T}^\eta d \eta_1 
\frac{h^+(\vec q,\eta_1) 
 S(\vec q, \eta_1) }{H^2 \eta_1^2},
\end{eqnarray}
with
\begin{equation}\label{defhpm}
 h^{\pm} ( q, \eta) = - \frac{H}{ \sqrt{2q} } \bigg( \eta \pm \frac{i}{q}\bigg) 
e^{\pm i q \eta}
\end{equation} 
and source
\begin{equation}
\label{S}
 S(\vec q, \eta) = \frac{16 \pi G H^2 \eta^2}{L^3} \sum_{\vec p} p_1 p_2 \bigg[ \bra \hat y (\vec q - \vec p) \ket \bra \hat 
y (\vec p) 
\ket \bigg]
\end{equation} 
($ y= a \delta \phi$ is the rescaled inflaton perturbation and $L$ is the size of a fiducial box in which the Fourier transform is performed). In these expresions we have omitted terms suppressed by the slow roll parameter $\epsilon$ (see \cite{big} for details).

As in the case of the scalar perturbations, in order to derive concrete predictions we need to consider an ensemble, 
calculate the stochastic CSL evolution of the inflaton field and compute the ensemble average. Shifting to the continuum (i.e., taking the limit $L\to \infty$) and using the results obtained in \cite{CSL-Inflation} for the expectation values in Eq. \ref{S}, for the tensor power spectrum we find
\begin{eqnarray}
\label{PStensor6}
 P^C_h (q) &=& \frac{ \pi^2 G^2 H^4 
\lambda_{0}^2}{ q^3 2^5} \int_0^{\infty} dv \int_{|1-v|}^{|1+v|} du \nn
 & & \frac{[4v^2-(u^2-v^2-1)^2]^2}{uv 
 [u+(u^2+4(\lambda_{0}/q)^2)^{1/2}][v+(v^2+4(\lambda_{0}/q)^2 )^{1/2} ] } \nn
& & \int_{-qT}^0 dx_1 \int_{-qT}^0 dx_2 \frac{1}{\sqrt{x_1 x_2}} \nn
& & J_{3/2} (x_1) J_{3/2} (x_2) X_{FF} (u,v;x_1,x_2) ,
\end{eqnarray} 
where
\begin{equation}\nonumber
 X_{FF} (u,v;x_1,x_2) \simeq x_1^2 x_2^2 \frac{[(qu)^2 + S_{qu}^2] [(qv)^2 + S_{qv}^2]}{4 
q^6} 
 \times 
 \end{equation}
 \begin{equation}\label{XFFapp}
 \bigg[ q^2 T^2 
+ x_2 (2 q T + x_2) \Theta (x_1-x_2) 
 + x_1 (2 q T + x_1) \Theta 
(x_2-x_1) \bigg]
\end{equation}
with $\lambda_0$ the CSL parameter, $S_{qu}$ and $S_{qv}$ defined in \cite{CSL-Inflation}, $J_{3/2}$ the Bessel function of order 3/2 and $\Theta(x)$ the Heaviside step function (the superscript $C$ in $P^C_h$ stands for `collapse' scenario). 

The result of Eq. \ref{PStensor6} is formally divergent; however, as discussed in \cite{CSL-Inflation}, one must introduce a physical cut-off $p_{UV}$. if we were interested in the tensor spectrum at the end of inflation we would take $p_{UV}$ to be given by the last scale that exits the horizon during inflation. However, since we are interested in the tensor modes we might observe at the CMB, we have to take into account the fact that, in order to reach us, primordial gravity waves have to go through the plasma era, where they would be subject to plasma damping effects. It is reasonable, then, to take the effective value of the cutt-off to be given by the scale of diffusion or Silk damping (see, e.g., \cite{damping}).

It is convenient to work with the dimensionless power spectrum defined as $\mathcal{P}^C_h \equiv q^3 P^C_h$. Doing so and keeping only the dominant terms, we finally arrive at
\begin{equation}\label{PStensor10}
 \mathcal{P}^C_h (q) \simeq 10^{-3} \frac{ G^2 H^4 \lambda_{0}^2 T^4 p_{UV}^5}{ q^3 } .
\end{equation} 
We note that the $1/q^3$ dependence implies that the best hope for seeing tensor modes would be to search for them at the largest possible scales. We also recall that the (dimensionless) scalar power spectrum found in \cite{CSL-Inflation} is given by
\begin{equation}
 \mathcal{P}^C_s (q) \simeq \frac{ G H^2 \lambda_{0} T}{\epsilon} ,
\end{equation} 
so the tensor spectrum can be written as
\begin{equation}
 \mathcal{P}^C_h (q) \simeq 10^{-3} \epsilon^2 \frac{ T^2 p_{UV}^5}{ q^3 } \left( \mathcal{P}^C_s (q) \right)^2 .
\end{equation} 
Therefore, in contrast to the linear relation obtained within the standard scheme (see Eq. \ref{usual-exp}), we obtain an expression for $\mathcal{P}^C_h$ which is quadratic in $\mathcal{P}^C_s$.

In order to estimate the magnitude of $\mathcal{P}^C_h$, we note that at the pivot scale of $q* = 0.05$ Mpc$^{-1}$ used by the Planck mission in order to measure the scalar perturbations, it was found that $\mathcal{P}^C_s = 10^{-9}$. As for the numerical value of $T$, we note that it depends on the energy scale of inflation and the total number of e-folds. If, as in \cite{CSL-Inflation}, we assume that the number of e-folds is 60 and that the characteristic energy scale of inflation is approximately $10^{-4}$ $M_P$, we obtain $T = 10^4$ Mpc. Finally, taking the two-fluid approximation of Seljak \cite{Seljak,silk1}, 
the Silk damping scale is $p_{UV}= 0.078$ Mpc$^{-1}$. Putting everything together we arrive at
\begin{equation}
 \mathcal{P}^C_h (0.05) \simeq 10^{-16} \epsilon^2.
\end{equation} 
Given that $\epsilon \ll 1$, our prediction for $\mathcal{P}^C_h$ renders it practically undetectable by current experiments. We can also estimate the tensor-to-scalar ratio within our model to get $r^C_{0.05} \simeq 10^{-7} \epsilon^2$, a value significantly smaller than the standard one, which is proportional to $\epsilon$ (see again Eq. \ref{usual-exp}).

We conclude that the failure to detect B-modes in the CMB should not be used to constrain the set of viable inflationary models. It is clear that the specific value obtained in our predictions above depends on some of the particularities of the inflationary model and on the UV cut-off considered. Nevertheless, our results are rather robust as these modifications would not change the fact that the prediction for the tensor spectrum depends quadratically on the scalar one and that it is suppressed by an extra factor of $\epsilon$.


It is worth mentioning that, while our analysis above is developed in a semi-classical setting, comparable modifications in estimates for the tensor power spectrum might hold in a wider class of scenarios. That is, even if one quantizes both the inflaton and metric perturbations, and not just the inflaton one as we did here, the current approach leaves room for predictions that differ dramatically from those of the standard account. This is because it is far from clear that the spontaneous collapses we employ to break the initial symmetries must affect in the same fashion the matter fields and the geometrical variables. In fact, it is reasonable to consider schemes where the collapse rates for matter fields and geometrical variables are different, or even schemes where the metric perturbations, although quantized, do not undergo spontaneous collapse by themselves.

The main conclusion we draw from the present analysis is that the prevailing view, which holds that the failure to detect primordial tensor modes can be used to rule-out specific inflationary models, is unjustified. The inability of the standard approach to account for the emergence of primordial inhomogeneities constitutes a serious drawback; it seems therefore rather hasty to relay on such a scheme in order to discard specific inflationary models, which, as we have shown, are still entirely viable within approaches designed to address such a shortcoming of the standard treatment. A farther reaching lesson we can extract from this work is that, at least in applications to cosmology, interpretational considerations within quantum theory can lead to dramatic modifications regarding observational issues. This serves to oppose a widespread attitude which regards such questions as of mere philosophical interest and dismisses their relevance regarding possible physical implications.

{\bf Acknowledgments } 
We acknowledge partial financial support from DGAPA-UNAM project IG100316 (EO and DS) and by CONACyT project 101712 (DS). GL's research funded by Consejo Nacional de Investigaciones Cient\'ificas y T\'ecnicas, Argentina. 
\bibliographystyle{apsrev4-1}
\bibliography{B-modes}

\end{document}